\documentclass[10pt,preprint]{aastex}

\graphicspath{{/Users/bahicks/Desktop/figures/}}

\usepackage{amsmath}
\usepackage{textcomp}

\shorttitle{Phase-Occultation Nulling Coronagraphy}
\shortauthors{Lyon, et al.}

\begin{document}

\title{Phase-Occultation Nulling Coronagraphy}

\author{Richard G. Lyon, Brian A. Hicks\altaffilmark{1}, Mark Clampin, Peter Petrone III}
\affil{NASA Goddard Space Flight Center, Greenbelt, MD 20771}
\email{richard.g.lyon@nasa.gov}

\altaffiltext{1}{NASA Postdoctoral Program Fellow. The NPP is administered by Oak Ridge Associated Universities.}


\begin{abstract}
The search for life via characterization of earth-like planets in the habitable zone is one of the key scientific objectives in Astronomy. We describe a new phase-occulting (PO) interferometric nulling coronagraphy (NC) approach. The PO-NC approach employs beamwalk and freeform optical surfaces internal to the interferometer cavity to introduce a radially dependent plate scale difference between each interferometer arm (optical path) that nulls the central star at high contrast while transmitting the off-axis field. The design is readily implemented on segmented-mirror telescope architectures, utilizing a single nulling interferometer to achieve high throughput, a small inner working angle (IWA), sixth-order or higher starlight suppression, and full off-axis discovery space, a combination of features that other coronagraph designs generally must trade. Unlike previous NC approaches, the PO-NC approach does not require pupil shearing; this increases throughput and renders it less sensitive to on-axis common-mode telescope errors, permitting relief of the observatory stability required to achieve contrast levels of $\leq10^{-10}$. Observatory operations are also simplified by removing the need for multiple telescope rolls and shears to construct a high contrast image. The design goals for a PO nuller are similar to other coronagraphs intended for direct detection of habitable zone (HZ) exoEarth signal: contrasts on the order of $10^{-10}$ at an IWA of $\leq3\lambda/D$ over $\geq10$\% bandpass with a large ($>10$~m)  segmented aperture space-telescope operating in visible and near infrared bands. This work presents an introduction to the PO nulling coronagraphy approach based on its Visible Nulling Coronagraph (VNC) heritage and relation to the radial shearing interferometer.
\end{abstract}

\keywords{Stellar coronagraphy, nulling interferometry, freeform optics, exoplanets, debris disks}

\section{Introduction}
\label{sec:intro}
For more than a decade, coronagraphic development efforts have been directed towards meeting the challenge of imaging and spectroscopic characterization of exoEarths within the habitable zone (HZ) of nearby stars. The goal is the detection of  spectral biosignatures that would provide evidence for the presence of life. Space telescopes will be required to achieve $10^{-10}$ contrast, since ground-based telescopes are ultimately limited by residual errors from the mitigation of atmospheric turbulence via adaptive optics. 

The Kepler mission has to date provided the best measurement of the fraction of stars having Earth-sized planets in the habitable zone, i.e. $\eta_{\earth}$. \cite{2014PNAS..11112647B} summarizes recent determinations of Kepler data suggesting that ``every late-type main-sequence star has at least one planet (of any size), that one in six has an Earth-size planet within a Mercury-like orbit, and that small HZ planets around M dwarfs abound,'' further noting Kepler data suggesting a potentially habitable planet residing within 5 parsec at the 95\% confidence level. While the actual value of $\eta_{\earth}$ remains unknown, \cite{2014ApJ...795..122S} adopt a baseline value of 0.1 while 0.2 is used by \cite{2015ApJ...799...87B} to estimate the number of stars that will need to be observed to successfully confirm detection of an exoEarth around a nearby star. \cite{2014ApJ...795..122S} show that characterizing dozens of exoEarths within a 5-year mission to obtain a meaningful statistical result for the presence of biosignatures will require apertures $>10$ meters in diameter. This result implies a segmented architecture coupled with a coronagraph be capable of working at $\leq 10^{−10}$ contrast at inner working angles of $\sim 2\lambda/D$.

The phase-occultation (PO) nulling approach described in this work differs from the existing Visible Nulling Coronagraph (VNC) described by \cite{2012SPIE.8442E..08L} in that it achieves increased throughput, discovery space, and starlight suppression (see Appendix \ref{appendix:lateral} for the basic principle of operation and additional background on the VNC). Advantages offered by the PO-NC include: (1) full-field observing in a single pointing - providing full planetary system snapshots every visit, (2) coronagraph throughput approaching 50\%, and (3) high-order stellar disk suppression with a single nulling interferometer (nuller), (4) high-contrast performance on a segmented aperture, at small IWA, and (5) insensitivity to telescope instability. Improved sensitivity derives simply from the fact that high-order stellar disk suppression for the PO-NC is delivered by a single nuller, thus significantly reducing the number of optics. A metric for optimization is contrast versus field angle targeting an inner working angle (IWA) $\leq 3\lambda/D$ where $\lambda$ is the central wavelength and $D$ is the aperture diameter.

There are additional benefits of the PO-NC that carry over from its VNC precursor, one of which being that it makes use of all available output photons for null control. This is due to the flux from the target star being conserved in the bright and dark output channels. This conservation law yields a robust null control approach that is independent of the state of instrument and its control temporal bandwidth depends only on the brightness of the target star. It does not levy beyond state of the art stability requirements on the telescope since the telescope must be stable only over each integration window for null control. The PO-NC design can also be implemented with reduced mass, volume, and integration complexity, with lower cost and risk.

\section{Mathematical formalism}
\label{sec:formalism}
\begin{figure}[t]
\begin{center}
\includegraphics[width=7cm]{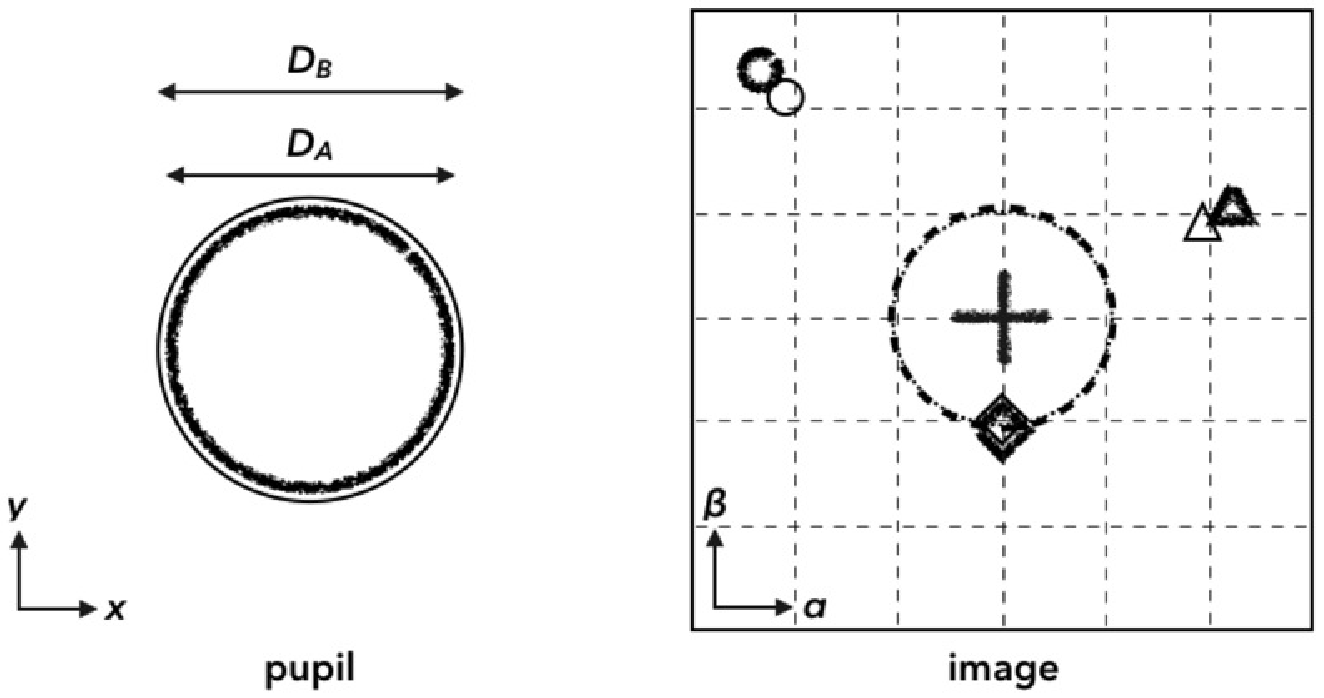}\\
\includegraphics[width=7cm]{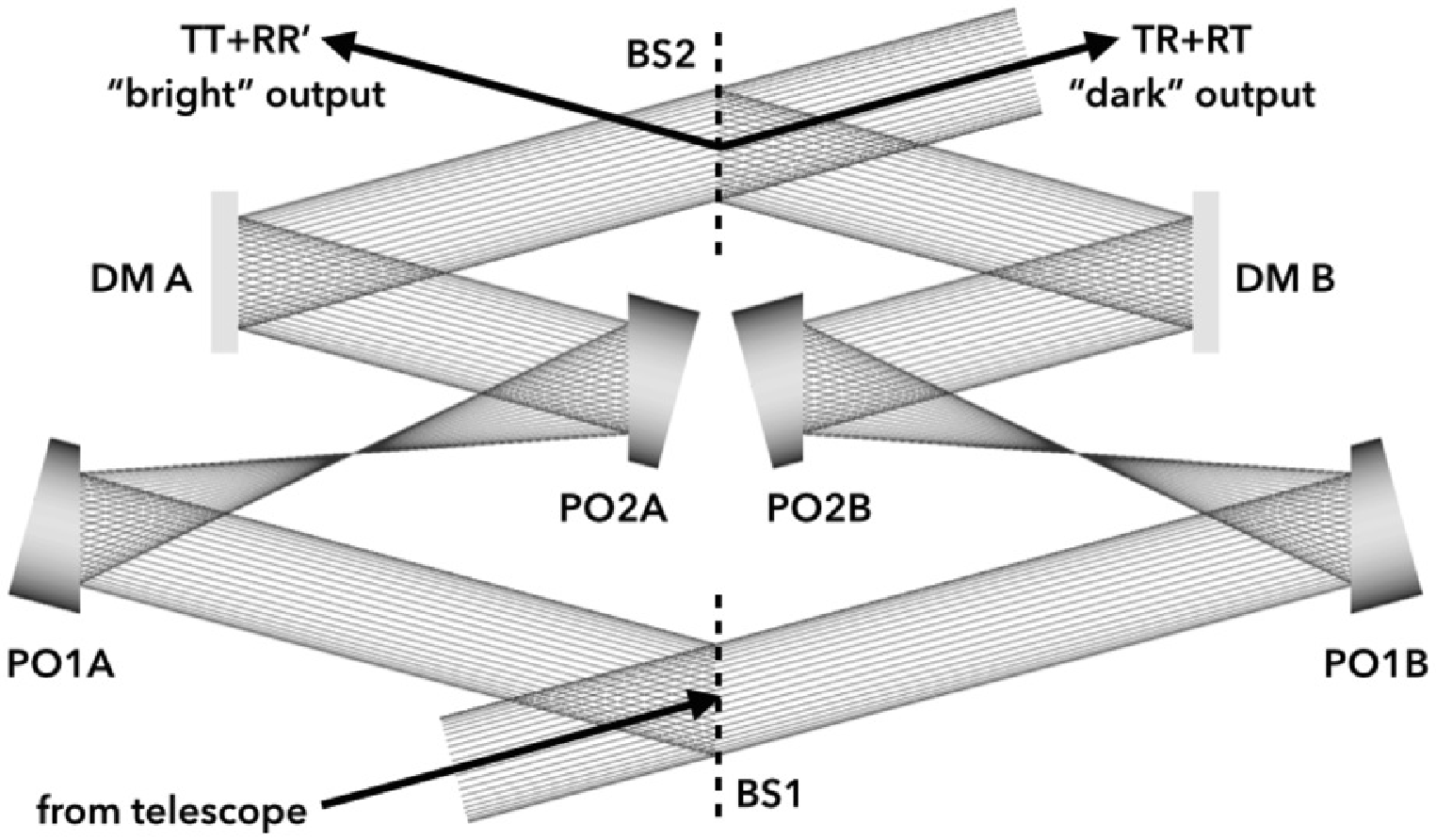}
\caption{A schematic phase-occulting (PO) nulling coronagraph layout including a deformable mirror (DM) in each arm to enable complex control of the full off-axis field of view. T denotes transmission and R denotes reflection through each beamsplitter (BS). The PO optics are OAPs with differing higher order aspheric terms in the ``A'' arm (heavy rough lines in the pupil and image footprint insets) and the relay in the  ``B'' arm (fine solid lines). The dashed lines indicate the approximate Airy disk (2.44$\lambda/D$) size for each arm, and each symbol is a different source center, with the cross being that of the central nulled star. }
\label{fig:schematic}
\end{center}
\end{figure}

The PO nulling coronagraph approach, shown in Figure \ref{fig:schematic}, uses a beamsplitter (BS1) to split the light from the telescopes exit pupil (lower left input) into left and right optical paths known as interferometer arms. The telescopes exit pupil is prior to BS1. The light in each arm is separately reflected off two-phase occulting optics, each with a focus between them and then reflected off a deformable mirror (DM) in each arm. Each DM is at a pupil and is an image of the telescopes exit pupil as relayed by the PO optics. The underlying surface shape of the PO optics are off-axis parabolas (OAP) but with added higher order aspheric terms such that the small beamwalk at each, since they are non-pupil optics, introduces differential optical distortion between the two arms. The differential distortion manifests itself as a small plate scale change between the two arms and combined with an overall phase shift due to the higher order aspheric terms introduces sixth-order focal plane Interferometric nulling. The second beamsplitter (BS2) acts as a beam combiner and combines the TR+RT into the beam exiting the PO-NC to the upper right, known as the ``dark'' output, and combines the TT+RR’ in the beam exiting to the upper left, known as the ``bright'' output. The dark output is brought to focus and used as the science channel while the bright output is used as a pointing and wavefront control channel.

The PO-NC’s differential distortion combined with the overall phase shift between arms achieves all of the desired performance criteria of symmetry, small IWA, high Strehl, and lower sensitivity to stellar angular extent. A field-dependent pupil phase \textit{difference} profile between interfering beams is given by
\begin{equation}
\phi(\mathbf{r},\boldsymbol{\alpha})=\phi_A(\mathbf{r},\boldsymbol{\alpha})-\phi_B(\mathbf{r},\boldsymbol{\alpha})
\end{equation}
where A and B refer to the left and right arms respectively, and where $\mathbf{r}=(x,y)=(r \cos \theta, r \sin \theta)$ are pupil coordinates normal to the beam propagation direction, normalized to a range of $[-\frac{D}{2},\frac{D}{2}]$, and $\boldsymbol{\alpha}=(\alpha, \beta)=(\rho \cos \gamma,\rho \cos \gamma)$ are Cartesian and polar angular coordinates of the object (on the sky) in units of $\lambda/D$. In the following we emphasize the azimuthal symmetry of the PO approach and define 
\begin{mathletters}
\begin{align}
r&=|\mathbf{r}|=\sqrt{x^2+y^2} \\
\rho&=|\boldsymbol{\alpha}|=\sqrt{\alpha^2+\beta^2}
\end{align}
\end{mathletters}

A phase profile that meets the desired performance characteristics can be expressed as
\begin{equation}
\label{eqn:phaseconditions}
\phi(r,\rho) =
  \begin{cases}
   \xi_1 & \text{if } 0\ \leq \rho < \epsilon IWA \\
   \xi_2 & \text{if } \epsilon IWA \leq \rho < IWA \\ 
   \xi_3 & \text{otherwise}
  \end{cases}
  \end{equation}
where
\begin{mathletters}
\begin{align}
\xi_1 & \equiv 0 \\
\xi_2 & \equiv\frac{2\pi s}{\lambda}P\frac{2r}{D}\cos(\theta-\gamma) \\
\xi_3 & \equiv \frac{2\pi s}{\lambda}\frac{2r}{D}\cos(\theta-\gamma)
\end{align}
\end{mathletters}
and
\begin{equation}
P=\Big(\frac{\rho-\epsilon IWA}{IWA(1-\epsilon)}\Big)^3
\end{equation}
and $s$ is a scale parameter that sets the coronagraphic IWA and $\epsilon$ extends from $[0,1]$ such that $\epsilon IWA$ is the radius of the ``flat spot,'' i.e. the central region of the phase where it is zero. If we look at the boundaries of the middle term in Equation \ref{eqn:phaseconditions} by setting $\rho=\epsilon IWA$ and $\rho=IWA$ to obtain $0$ and $\frac{2 \pi s}{\lambda}\frac{2r}{D}\cos(\theta-\gamma)$, respectively, these are the same as the first and third terms in Equation \ref{eqn:phaseconditions} ensuring continuity. The middle term in Equation \ref{eqn:phaseconditions} takes the form of optical distortion, i.e. field-dependent magnification that varies as the cube of the field angle, which results in a field-dependent plate scale. It is seen from Equation \ref{eqn:phaseconditions} that the on-axis phase difference is zero and that the off-axis optical path difference between the two arms increases as the cube out to the IWA and remains constant for field angles outside the IWA. In practice the on-axis phase difference between the two arms is set to $\pi$ by increasing the optical path length of one arm relative to the other by $\lambda/2$. This sets the on-axis phase difference to $\pi$ to meet the condition of central nulling and the off-axis phase difference decreases to zero outside the IWA to allow off-axis planet light to pass through.

A schematic of a phase-occulting (PO) nulling coronagraph layout with corresponding pupil and image coordinate definitions are shown in Figure \ref{fig:schematic}. Slight differences in field dependent angular magnification (i.e. distortion) and corresponding areal beam compression conserve etendue between each arm of the nuller and this is exaggerated in the pupil and roughly to scale in the image. There are significant platescale differences between beams near the IWA such that structure including a tilt offset is deliberately added to the wavefront in each arm, making it such that the imaged point spread function (PSF) associated with each beam do not exactly overlap and do not destructively interfere. Beyond the IWA, the PSFs associated with each beam no longer separate, thereby maintaining a Strehl ratio that is comparable to what would be achieved for a single diffraction-limited optical train.

A key difference between the PO optics and what would appear to be off-axis-parabola (OAP) mirrors used in Figure \ref{fig:schematic} is that the on-axis regions of the PO optics are relatively flat such that there is little sensitivity to the finite angular extent of stars, or similarly, the residual pointing jitter of the telescope that is stabilized prior to entering the nuller that blurs as a probability distribution of zero mean about the true center of the star (see Section \ref{sec:prospective}). For an illustrative example of a related but lesser-performing system, the reader is referred to Appendix \ref{appendix:radial} describing a radial shearing inteferometer used as a nuller that introduces a constant difference in plate scale, or equivalently, magnification between beams.

Taking $\sqrt{R}$, $\sqrt{R'}$, and $\sqrt{T}$ as the complex reflectivities (prime denotes reverse direction) and transmissivity of each  beamsplitter, for an ideal lossless beamsplitter the phase difference between each reflected and transmitted beams is $\pi/2$. This is ascribed to the reflectivities such that $\sqrt{R(')}\rightarrow \sqrt{R(')} e^{i\pi/2}$. In order to minimize the departure from parabolic curvature for each optic, oppositely signed higher order aspheric terms are applied to the PO optics in each arm such that the electric fields corresponding to the four paths that follows for the intermediate field angles (the second condition in Equation \ref{eqn:phaseconditions}) can be written as
\begin{mathletters}
\begin{align}
\label{eqn:fourbeams}
\psi_{RR'} & = \sqrt{RR'} e^{i\pi} e^{i2\pi(\alpha x + \beta y)}e^{i\xi_2} \\
\psi_{RT} & = \sqrt{RT}e^{i\pi/2} e^{i2\pi(\alpha x + \beta y)}e^{i\xi_2} \\
\psi_{TT} & = \sqrt{TT} e^{i2\pi(\alpha x + \beta y)}e^{-i\xi_2}e^{i\pi} \\
\psi_{TR} & = \sqrt{TR}e^{i\pi/2} e^{i2\pi(\alpha x + \beta y)}e^{-i\xi_2}e^{i\pi}
\end{align}
\end{mathletters}
The two fields that transmit through the first beamsplitter $(\psi_{TR},\psi_{TT})$ have an additional phase shift of $\pi$ accomplished by delaying the ``B'' arm by a half wave broadband to facilitate nulling in the symmetric output. The fields in the bright and dark outputs are given by the sums
\begin{mathletters}
\begin{align}
\psi_{bright}&=\psi_{RR'}+\psi_{TT} \\
\psi_{dark}&=\psi_{RT}+\psi_{TR}
\end{align}
\end{mathletters}
Simplifying to the case of an ideal 50/50 beamsplitter such that $\sqrt{RR'}=\sqrt{TT}=\sqrt{RT}=\sqrt{TR}=1/2$ and applying trigonometric identities the bright and dark output fields may be expressed as 
\begin{mathletters}
\begin{align}
\psi_{bright}&=e^{i2\pi(\alpha x + \beta y)}e^{i\pi}\cos{\xi_2}  \\
\psi_{dark}&=ie^{i2\pi(\alpha x + \beta y)}e^{i\pi/2}\sin{\xi_2}
\end{align}
\end{mathletters}
The pupil plane intensities in the two outputs are then
\begin{mathletters}
\begin{align}
I_{bright}&=\cos ^2 \Big[\frac{2\pi s}{\lambda}P\frac{2r}{D}\cos(\theta-\gamma)\Big]  \\
I_{dark}&=\sin ^2 \Big[\frac{2\pi s}{\lambda}P\frac{2r}{D}\cos(\theta-\gamma)\Big]
\end{align}
\end{mathletters}

Integrating the bright and dark intensities over the area of a circular exit pupil yields the bright and dark output transmission as
\begin{mathletters}
\begin{align}
\label{eqn:brighttransmission}
T_{bright}(\rho)&=\frac{1}{2}+\frac{J_1\Big(\frac{4\pi s}{\lambda}P\Big)}{\frac{4\pi s}{\lambda}P}  \\
\label{eqn:darktransmission}
T_{dark}(\rho)&=\frac{1}{2}-\frac{J_1\Big(\frac{4\pi s}{\lambda}P\Big)}{\frac{4\pi s}{\lambda}P}.
\end{align}
\end{mathletters}

\begin{figure}[t]
\begin{center}
\includegraphics[width=5cm]{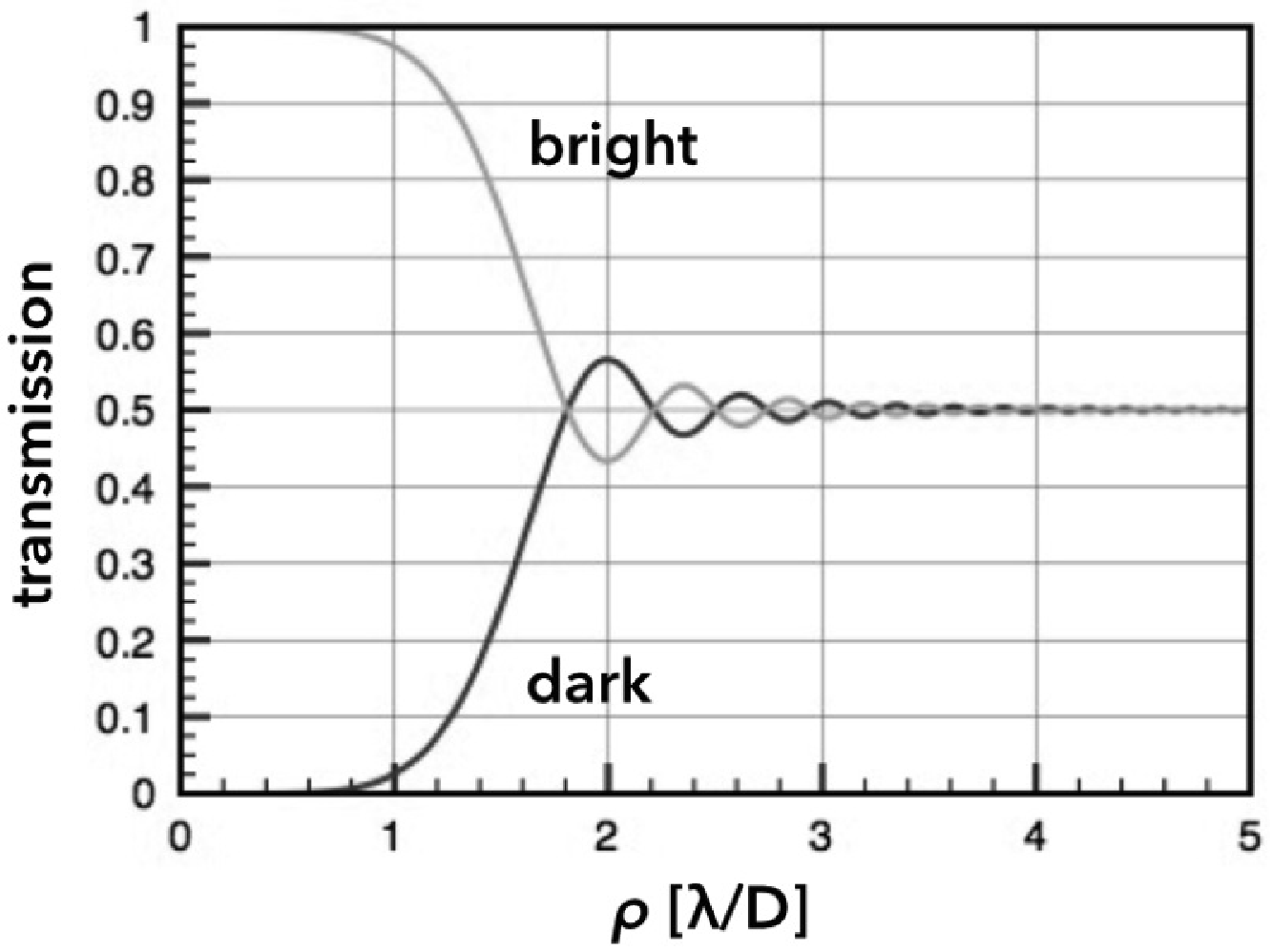} \\
\includegraphics[width=5cm]{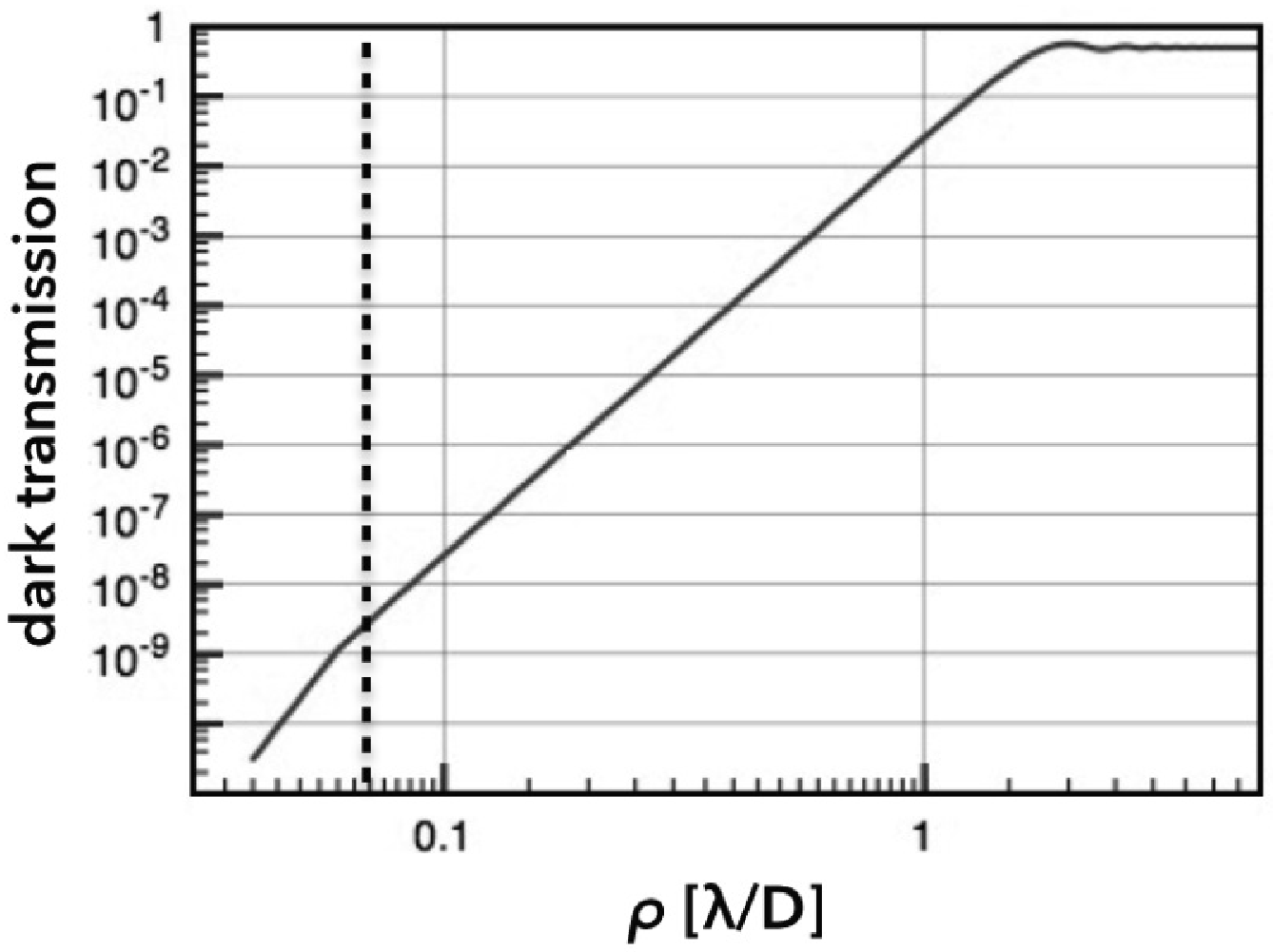}
\end{center}
\caption{\textit{Top}: A linear plot of PO nuller targeted bright (TT+RR') and dark (TR+RT) output transmission profiles. The dark science output transitions to transmission at approximately $2\lambda/D$. Energy conservation is satisfied by the profiles summing to unity at all points. \textit{Bottom}: A log-log plot of the transmission profile that scales as $\rho^6$ for separations less than the $2\lambda/D$ IWA. The vertical dashed line marks the outer edge of a G2V star observed at 10~pc with a 10~m telescope operating at 500~nm.}
\label{fig:transmission}
\end{figure}

The bright and dark throughputs are plotted versus $\rho$ in the top of Figure \ref{fig:transmission} and the dark throughput is plotted on a log-log scale in the bottom panel of Figure \ref{fig:transmission}, both of which assume $\epsilon=0$ and $IWA=2\lambda/D$ and s is set such that the first transmission peak is at the IWA. The log-log plot shows the $\rho^6$ dependence of contrast versus field angle.

The modulus squared of the Fourier transform of the complex dark pupil produces the image plane point spread function (PSF) that is the modulus square of the complex sum of the interfering beams with an intensity that varies as $\rho^6$ for small angles and the integrated leakage due to the finite extent of the stellar disk scales proportionately as shown in Figure \ref{fig:transmission}. The Fourier transform of the complex dark pupil is denoted as the amplitude spread function (ASF) and is shown here as:
\begin{eqnarray}
\label{eqn:asf}
ASF(\eta_x,\eta_y)=\frac{1}{2}e^{i\pi/2}\tilde{A}(\eta_x,\eta_y)** \nonumber \\
\big[\delta(\eta_x-\alpha-\zeta_x)\delta(\eta_y-\beta-\zeta_y)- \nonumber \\
\delta(\eta_x-\alpha+\zeta_x)\delta(\eta_y-\beta+\zeta_y)\big]
\end{eqnarray}
where $(\eta_x,\eta_y)$ are the focal plane angular coordinates projected on the sky and ``$**$'' denotes a 2D convolution over $(\eta_x,\eta_y)$, $\delta(...)$ signifies the Dirac delta function, $\tilde{A}(\eta_x,\eta_y)$ is the Fourier transform of the pupil shape function (aperture), and
\begin{mathletters}
\begin{align}
\zeta_x&=\frac{2 s}{\lambda}P\cos\gamma \\
\zeta_y&=\frac{2 s}{\lambda}P\sin\gamma
\end{align}
\end{mathletters}
are the distortion shifts in focal plane angular coordinates, also in units of $\lambda/D$. It is seen from Equation \ref{eqn:asf} that the ASF is the difference of two shifted $\tilde{A}(\eta_x,\eta_y)$ located in the focal plane at coordinates
\begin{equation}
\label{eqn:asf2}
(\eta_x,\eta_y)=(\alpha,\beta)\pm\frac{2s}{\lambda}P(\cos\gamma,\sin\gamma)
\end{equation}

The PSF is given by $PSF=|ASF|^2$ and consists of two bright peaks with interference between them. It is seen from Equation \ref{eqn:asf2} that the center between the two peaks moves linearly in the focal plane for increasing $(\alpha,\beta)$ and with a separation of 
\begin{equation}
\Delta \rho = \frac{4s}{\lambda}P.
\end{equation}
Assuming the target planet is at the IWA yields $P=1$ and defining the IWA as the minimal angular separation at which the dark transmission (Equation \ref{eqn:darktransmission}) reaches 1/2, then the Bessel function term in Equation \ref{eqn:darktransmission} would be zero when Bessel function argument is $1.22\pi=4\pi s/\lambda$ or when $s=0.305\lambda$.

\begin{figure*}[h]
\begin{center}
\includegraphics[width=16cm]{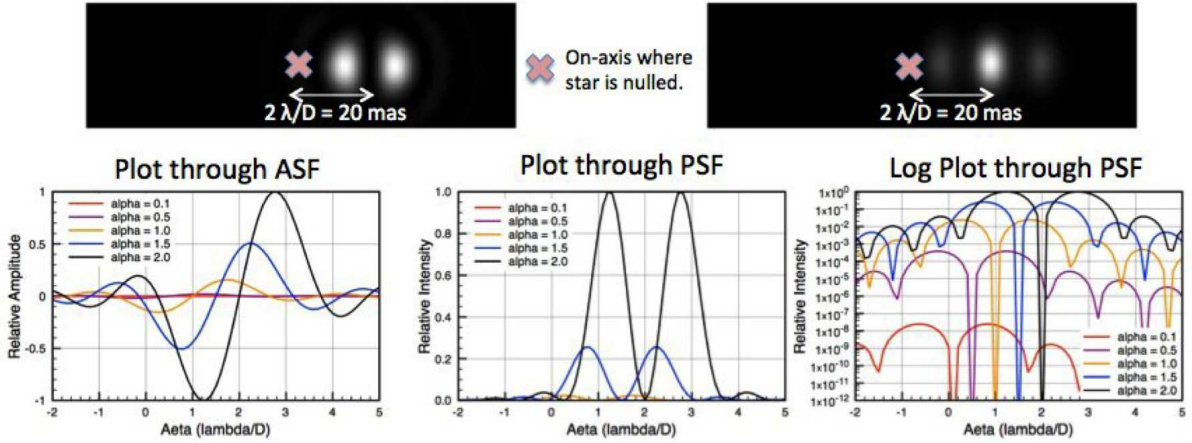} 
\end{center}
\caption{The PO-NC uses aspheric surfaces overlaid on a parabolic surface to introduce cubic field dependent phase shifts that results in the difference of two shifted ASFs. Their modulus squared gives the PSF response to a single off-axis planet (upper left). Two peaks are visible such that the half way point between them is at $2\lambda/D$. A simple matched filter (photon-noise limited) is used to sum the flux into the correct location at $2\lambda/D$ (upper right). The lower row plots the ASF, PSF and log-PSF versus the focal plane coordinate for left to right in the upper left panel. Plots are shown for planet locations of $\boldsymbol\alpha= 0.1, 0.5, 1.0, 1.5, 2.0 \lambda/D$ to show the increased peak intensity that reaches its maximum at the $2\lambda/D$ IWA.}
\label{fig:ASF}
\end{figure*}

In order to further improve signal statistics, a matched filter can be used to merge the two peaks into one and this is shown schematically in Figure \ref{fig:ASF}. The theory of matched filters is well developed. Here we employ a version such that the filter coefficients sum to zero and the square of the filter coefficients sum to unity. The latter ensures that the counts within the peak of the matched filter output is the integral of the counts within the two shifted PSFs. The zero sum ensures that it sets background terms that are slowly varying to zero, with the filter being optimal for detection of point sources against a constant background. Some complexity arises in that the dual PSFs rotate about the center of the focal plane and the matched filter must therefore do the same.  The matched filter can fortunately be generated from Equation\ref{eqn:asf} with arbitrary rotations with the focal plane broken into azimuthal zones, one zone per rotation, over which each matched filter can be applied.

\section{Design considerations}
\label{sec:design}

A description of the target PO nulling performance has been given above, and here we emphasize some of the important practical aspects of the optical design including manufacture of surfaces with slight but significant departure from parabolic, pupil plane relay, and beamwalk, with the latter two being critical to the function of the PO principle of operation.



The conditions for phase-occultation asserted in Equation \ref{eqn:phaseconditions} added to e.g. off-axis parabola (OAP) surfaces to become PO optics that generate the progressive null to transmission profile specified in Equation \ref{eqn:fourbeams} imply zonal aspheric optical surface profiles. Generating such surfaces may be accomplished using computer controlled polishing techniques such as magnetorheological finishing (MRF), which has demonstrated capability of achieving $<1.5$~\AA \,roughness on fused silica as measured by atomic force microscopy \citep{2014SPIE.9281E..09D}. The precision to which the necessary freeform optics will need to be polished and aligned will scale with instrument size. It will be preferable to make these optics and the nuller itself as small as possible to minimize susceptibility to thermal excursions, but this implies more magnification downstream from the telescope which results in more demanding fine pointing system requirements. Similarly, a decrease in size means tighter alignment requirements. A complication that must be considered is that surface and alignment tolerancing and measurement will not necessarily follow standard practices since Zernike or Seidel aberrations measured by phase-shifting interferometers may not be used to characterize a single surface.

Achieving high-order suppression and small IWA while maintaining a reasonable imaging Strehl ratio (measured relative to a single beam) off-axis requires keeping the PSFs of each arm at a fixed narrow separation, which implies field-dependent distortion introduced by the pairs of freeform PO optics. We show a DM in each arm for correcting complex wavefront asymmetries due to misalignments and polishing defects, and these must lie in a pupil in order to correct the full field. The PO optics act not only as the centro-symmetric nulling optics, but also pupil relays. A design challenge is locating the pupils before and after the PO optics not only to ensure there is clearance for mounts between the folded beams, but also so that there is adequate beamwalk across each PO surface. In this sense Figure \ref{fig:schematic} is only a schematic, beam compression may be used between the PO1 and PO2 optics to step down from what is expected to be a fast steering mirror that is also in a pupil upstream from the nuller, but downstream from the telescope in order to stabilize body pointing jitter of the observatory attitude control system as described in Section \ref{sec:prospective}.

Mean DM motion corrects the piston difference between beams, i.e. monochromatic phase delay, and the higher order on the DM corrects the RMS WFE and can be used to achieve perfect monochromatic nulling on-axis. The key to broadband operation is to set zero path length deviation between both arms over the design bandpass then introduce the path length deviation as a variation about a mean of zero across the wavefront. This varies the size of the dark hole where dark hole size is inversely proportional to wavelength, i.e. the longest wavelength sets the IWA and the shortest sets the OWA. One way of making the PO optics work broadband is through the use of Fresnel rhombs to introduce a polarization-based $\pi$ phase shift that varies slowly with wavelength and is tunable to a given bandpass. The practical implementation of using such Fresnel rhombs as achromatic phase shifters is described in \cite{2014SPIE.9143E..2SH}.


\section{Prospective application to a large segmented space-based telescope}
\label{sec:prospective}

As suggested in Section \ref{sec:intro}, a 10-meter space telescope will be well-suited to direct imaging and characterization of exoplanets. Such a large aperture in space will achieve the requisite photometric sensitivity, resolution, and contrast for accessing the habitable zones around nearby stars to perform spectroscopy of Earth-like worlds. The fairing size of current and suggested launch vehicles puts a fundamental limit on the size of monolithic primary mirror apertures that can be launched. Even the largest planned fairing will only be able to house an 8-meter monolith. The likely path forward for achieving a space telescope with larger aperture sizes is to build on the experience of JWST using a deployable segmented aperture. A segmented mirror introduces a different starlight diffraction problem that few existing coronagraph designs can accommodate. A nuller such as the PO-NC, however, can be paired with not only a segmented aperture, but also the full range of types including unobscurred, obscured, and sparse or dilute aperture telescopes.

\begin{figure*}[t]
\begin{center}
\includegraphics[width=12cm]{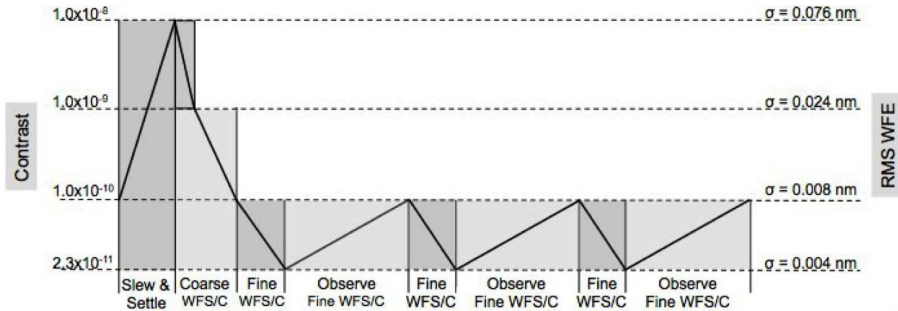}
\end{center}
\caption{The nominal observing sequence for the PO-NC. Instrument contrast is expected to degrade due to increased wavefront error (WFE) introduced by thermally induced disturbances during slewing. Following target acquisition, during the settle time, coarse WFS/C is initiated to bring the system to within range of reasonable contrast. Fine WFS/C further reduces WFE to picometers to achieve the contrast floor. During an observation fine WFS/C operates in closed-loop to hold the contrast at or below the $10^{-10}$ contrast floor suitable for exoEarths.}
\label{fig:observation_sequence}
\end{figure*}

Instabilities of both the telescope as a whole, the segments that comprise the primary aperture, and the entire optical train affect all coronagraphs. In addition to reaching the residual WFE shown in Figure \ref{fig:observation_sequence} through WFS/C, the ability of the fine pointing system (FPS) to reduce body pointing jitter in the telescope is crucial to minimize the leakage attributed to moving the star relative to the center of the coronagraphic transmission pattern, which may be thought of as an effective increase in the (unresolved) stellar angular diameter, and unwanted WFE attributed to beamwalk downstream from the FPS. The aforementioned VNC and also the Phase Induced Amplitude Apodization Complex Mask Coronagraph (PIAACMC) \citep{2014ApJ...780..171G} are two coronagraphs that can work with arbitrary apertures that have been modeled to reach contrasts of $4\times 10^{-8}$ at $3\lambda/D$  \citep{2014SPIE.9143E..0VK} in the presence of 1.6~mas pointing jitter expected for the planned WFIRST-coronagraph . The PIAACMC is a more conventional single optical train coronagraph susceptible to instabilities throughout the train, whereas the phase-occulting (PO) implementation of the VNC described in this work can be designed such that instabilities of large telescope optics traversing long focal lengths upstream of the interferometric cavity are seen as being common-mode, and it is only the instabilities internal to the interferometric cavity that dynamically compromise contrast. 

Beamwalk on all non-pupil optics prior to the first beamsplitter will introduce WFE from the telescope. The on-axis WFE occurs common-mode but because of the tip/tilt walking the WFE off-axis it is no longer true common mode. Taking into account jitter with sixth-order nulling and maximum transmission of $\sim50$\% occurring at an IWA of (2$\lambda/D$), for a systematic off pointing of $0.1\lambda/D$ the stellar leakage would be $0.5(0.1 / 2 )^6 = 7.81\times 10^{-9}$. In practice pointing jitter should manifest as a mean zero probability distribution. Assuming a circular Gaussian distribution with $1-\sigma$ being $0.1\lambda/D$, the integrated $\theta^6$ error is $\approx 3.75\times 10^{-7}$. This leakage in the focal plane would look like a central spot about 2 to 3 orders brighter than a $10^{-10}$ planet. Assuming a typical PSF profile to decrease in intensity as $\sim 1/(1 + \rho^3 \pi^4/8 )$,  there is $\sim100$ less leakage at  $2\lambda/D$ giving $\sim 3.75\times10^{-9}$ contrast. Assuming no other contributions to the error budget, to achieve $10^{-10}$ the residual pointing jitter would need to be $\leq0.055\lambda/D$ (1-$\sigma$), which for a 10-m telescope operating at $\lambda=500$~nm this implies a 1-$\sigma$ jitter of 0.57~mas.

The actual design will dictate what level of sensitivity can be achieved. Polishing a flat spot in the aspheric departure of the PO optics would mitigate this leakage as described in Section \ref{sec:formalism}. With the higher order aspheric departure starting at $\pm 0.5\lambda/D$, increased pointing stability out to of order $0.25\lambda/D$ would be achieved due to the non-shearing and non-distorting condition introduced by the flat spot. In other words, the WFE and pointing jitter would appear common mode on-axis out to $0.25\lambda/D$ as described in Equation \ref{eqn:phaseconditions}.

\begin{figure*}[t]
\begin{center}
\includegraphics[width=12cm]{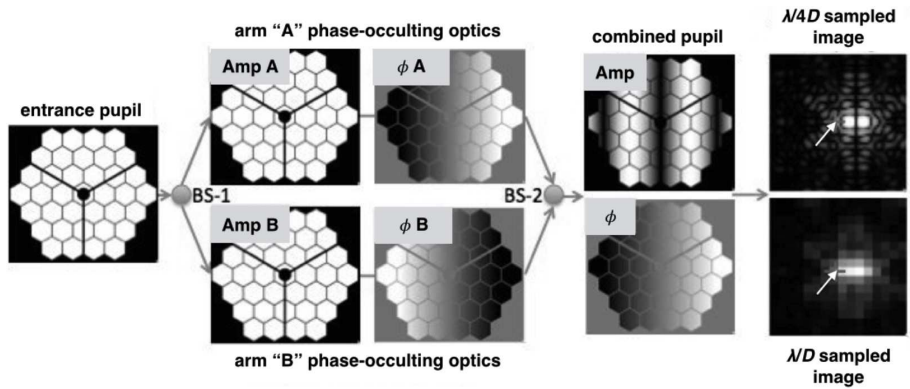}
\end{center}
\caption{A simulation of the PO optics used with a segmented aperture showing complex components of the electric field inside the interferometer and following combination in both pupil and image planes at $2\lambda/D$ off-axis. The dark detector focal plane images show critical and oversampling with white arrows indicating the location of the suppressed star.}
\label{fig:PO-segmented}
\end{figure*}

The geometry of the overlap between interfering beams determines both throughput and transmission on the sky. The PO nulling coronagraph benefits from needing only a very small amount of distortion to achieve separation between the beams associated with an off-axis source, so the entrance aperture very closely resembles the Lyot mask that is used to reject regions of non-overlap between interfering beams. An example entrance and (re)combined exit pupil pair is illustrated in Figure \ref{fig:PO-segmented} along with the basic optical principle of operation at $2\lambda/D$.  Following splitting and after reflecting off the PO optics as shown in Figure \ref{fig:schematic}, the beams have the amplitudes labeled ``Amp'', an aperture function equal to $\sqrt{R}$ and $\sqrt{T}$ for arms ``A'' and  ``B'' where non-zero, respectively, and phases $\phi$ A, B where the additional off-axis tilted phase that is common to both arms has been removed such that only aspheric departure is shown. This shape is approximately linear across the pupil with a slope that varies approximately as the cube of the star-to-planet angular separation. The combined amplitude is shown as its modulus and contains two cycles per aperture; the phase is primarily tilt corresponding to an inner working angle of $2\lambda/D$. The effect of this amplitude and phase is to produce two focal plane planet PSFs with a mean $\pi$-phase shift between them, but separated such that the center is at $2\lambda/D$. The PSFs are shown on a log-scale and sampled at $\lambda/(4D)$ to show detail and Nyquist-limited. The location of the nulled star is indicated by an arrow in each image.

In addition to meeting IWA and contrast requirements, it is necessary for a coronagraph design to achieve the highest possible throughput to maximize sensitivity. Coronagraph Lyot masks typically block a portion of the aperture thereby lowering useful aperture. Maximizing the useful aperture yields larger photon rates, which (i) provides increased sensitivity for achieving significant detections in images and faint spectroscopic signals; (ii) reduces observational overhead associated with initial wavefront sensing and control (WFS/C) cycle after slewing and settling on a new science target (see Figure \ref{fig:observation_sequence}); (iii) achieves higher closed-loop WFS/C bandwidth to mitigate telescope and coronagraph stability and drift requirements. WFS/C  drives important operational design considerations for a coronagraph since the coronagraph  has to be able to converge to the required contrast level and then maintain closed-loop control during the science exposure as illustrated in Figure \ref{fig:observation_sequence}. 

The initial high-contrast WFS/C portion of the observational duty cycle (i.e. the coronagraph overhead) scales inversely with instrument throughput, with higher throughput yielding higher WFS/C bandwidth \citep{2012OptEn..51a1002L}. During the lifetime of a mission, savings in initial setup overhead provides a significant boost in efficiency.  When considering telescope stability requirements, shorter WFS/C timescales (higher control bandwidth) allow greater telescope drift per unit time for the science observations and ensure that the WFS/C will actually be successful in converging the coronagraph to the operating contrast. Thus throughput can be a deciding factor in determining whether a faint target is inaccessible, accessible for direct imaging but not spectroscopy, or for both imaging and spectroscopic follow-up.

\begin{figure}[t]
\begin{center}
\includegraphics[width=7cm]{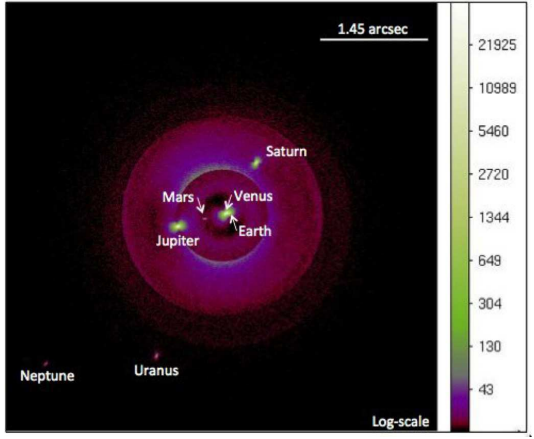}
\end{center}
\caption{Simulated PO-NC 30,000~s V-band image of the Solar System observed face-on at a distance of 10~pc with a 10~m diameter 37-segment aperture telescope. The image is photon-limited with the color bar in units of photo-electrons. The signal of a the stellar disk would leak at a level of $10^{-8}$ for $\sim 10^{-10}$ contrast at $2\lambda/D$ to be further reduced in post-processing. Here the stellar residual is removed altogether to emphasize the PSF of the match-filtered PO-NC and relative counts from planets, dust, and debris signals, the latter of which is a hypothetical distribution provided courtesy of Chris Stark.}
\label{fig:solarsystem}
\end{figure}

A simulation of the Solar System as observed face-on at 10~pc with a 36 segment hexagonal array telescope and the PO-NC is shown in Figure \ref{fig:solarsystem} emphasizing the PSF of unresolved and extended objects in the field. The telescope and coronagraph are assumed to be ideal with no pointing jitter and perfectly stable optics. The field of view of a coronagraph would be much smaller. The number of counts assume an 30,000 s (8 h 20 m) exposure.

\section{Summary}
\label{sec:summary}

We have described a nulling coronagraph based on PO optics. The PO coronagraph is a new, enhanced nulling design that needs only a single interferometer to introduce small amounts of distortion in each arm to meet all coronagraphic performance goals for direct detection of exoplanetary signal. It offers several key benefits for future large aperture space telescopes that would seek to image earths. Significantly improved throughput yields a sensitivity gain and increases the efficiency of WFS/C for convergence to the required contrast with increased control bandwidth. Full-field transmission removes the need to build an image from a sequence of varied shears and/or telescope rotations, so observational operations are significantly simplified. One-to-one on-axis pupil mapping also mitigates instability in the telescope primary mirror thereby relaxing telescope requirements.

The PO-NC design makes use of freeform optical surfaces to achieve high-order starlight suppression and full off-axis discovery space. We have shown that the  PO-NC concept has the potential  to undertake direct imaging and spectroscopic characterization of extrasolar terrestrial planets in the habitable zone by achieving contrasts on the order of $10^{-10}$ at an inner working angle (IWA) of $\leq3\lambda/D$ with, e.g. a future large aperture ($\gtrsim10$-m) segmented space-based telescope.

\acknowledgments
The authors acknowledge support from NASA/Goddard Space Flight Center Internal Research and Development (IRAD), and from the NASA Strategic Astrophysics Technology (SAT) Technology Development for Exoplanet Missions (TDEM) program.

\appendix 
\section{The lateral shearing Visible Nulling Coronagraph}
\label{appendix:lateral}
\begin{figure}[t]
\begin{center}
\includegraphics[height=3.5cm]{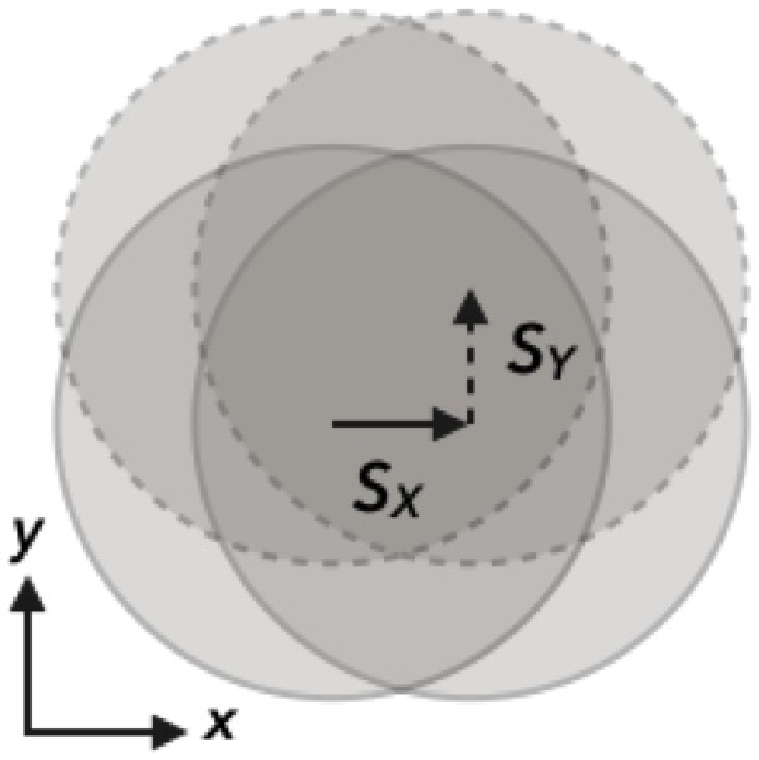} 
\includegraphics[height=3.5cm]{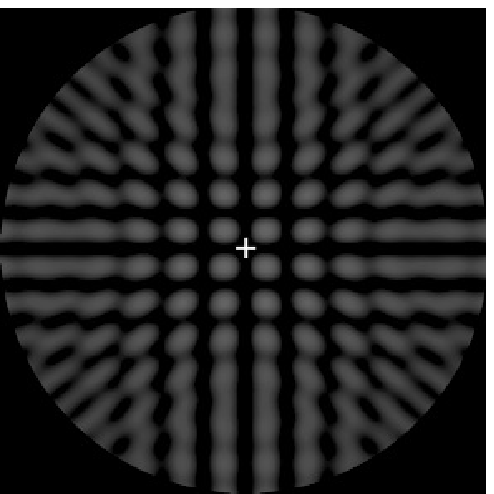} \\
\includegraphics[width=7cm]{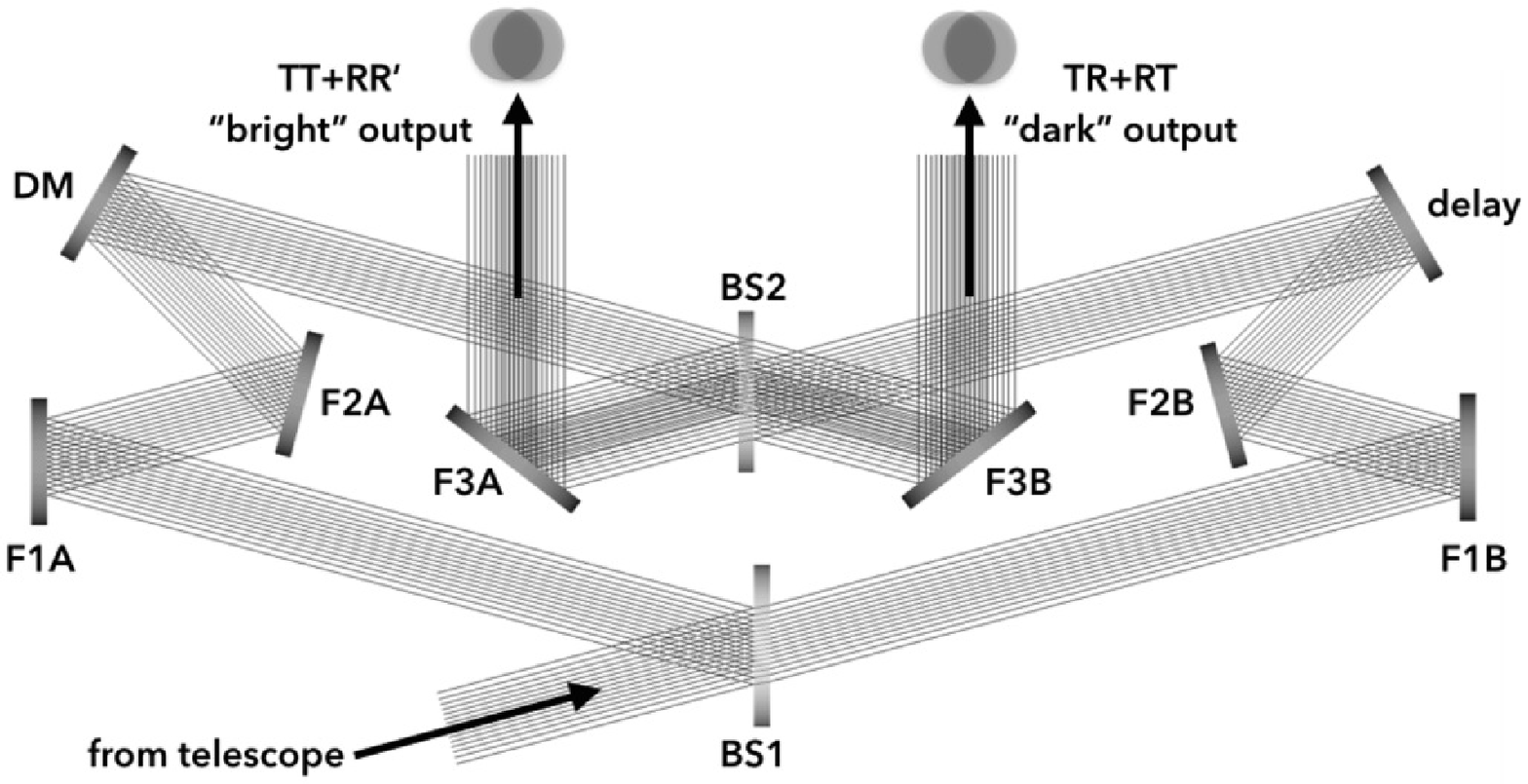}
\end{center}
\caption{A single stage of the lateral shearing VNC with all fold (F) mirrors. The dark output of this first x shear stage would be the input to the second stage that shears in y as shown in the upper left. The region where all four beams overlap is the effective area, all other non-overlapping regions are masked out. The upper right shows polychromatic on-sky transmission of the dark output - the cross marks the location of the nulled star.}
\label{fig:lateral}
\end{figure}

The VNC is presented in detail through several works and has been proposed for large-scale missions including Extrasolar Planetary Imaging Coronagraph (EPIC) described in \citep{2004SPIE.5487.1538C} and \cite{2008SPIE.7010E..45L}. Here we briefly summarize for convenience the basic principle of operation and theoretical performance. Laboratory-based results to date \citep{2009SPIE.7440E..11L, 2012SPIE.8442E..08L} have achieved $5.7\times10^{-9}$ contrast over a 1~nm wide band centered at 633~nm. It has been the decade-long development of the lateral shearing VNC schematized in Figure \ref{fig:lateral} that has served to motivate the PO optics described in this work.

Key advantages of the lateral shearing VNC approach that are common with the PO-NC include: 1) compatibility spanning all future flight telescope architectures including unobscured, obscured, segmented, and sparse apertures, 2) access to all starlight photons for WFS/C, and 3) a small IWA. One of the complications associated with the VNC is the need for two nullers in order to achieve fourth-order suppression of a star's unresolved disk (a single nuller is only second-order), which is most problematic for nearby stars for which photometric detection of exoplanets and debris disks is most favorable.

An added disadvantage of the laterally sheared VNC in one or two shears which may be in the same direction or orthogonal is the periodic on-sky transmission pattern shown in Figure \ref{fig:lateral}, which may be expressed as
\begin{equation}
T(\theta_x,\theta_y)=\sin^2\Big(\pi \frac{s_xD}{\lambda}\alpha\Big)\sin^2\Big(\pi \frac{s_yD}{\lambda}\beta\Big)
\end{equation}
where $S_x=s_xD$ and $S_x=s_yD$ are shears related to fractional aperture shears, not to be confused with the scale factor $s$ defined in Section \ref{sec:formalism}. In order to fill in the full sky where at a given epoch an exoplanet may be located in a region of destructive interference, either the telescope is rolled or the shear is varied or a combination of both is used in combination. Both shearing and rolling imply additional time needed for wavefront control for any one configuration. Rolling on a target adds associated slew and settle overhead and a shear stage in one or both nullers introduces mechanisms. While neither the need to roll or shear represents a prohibitive requirement, shifting the design complication to the design, fabrication, and alignment of the phase-occulting optics should in theory produce a more elegant approach that achieves full discovery space, high-order stellar disk suppression, and high throughput with a single nuller with no moving parts.

\section{A basic radial shearing interfometer as a nuller}
\label{appendix:radial}

\begin{figure}[t]
\begin{center}
\includegraphics[width=7cm]{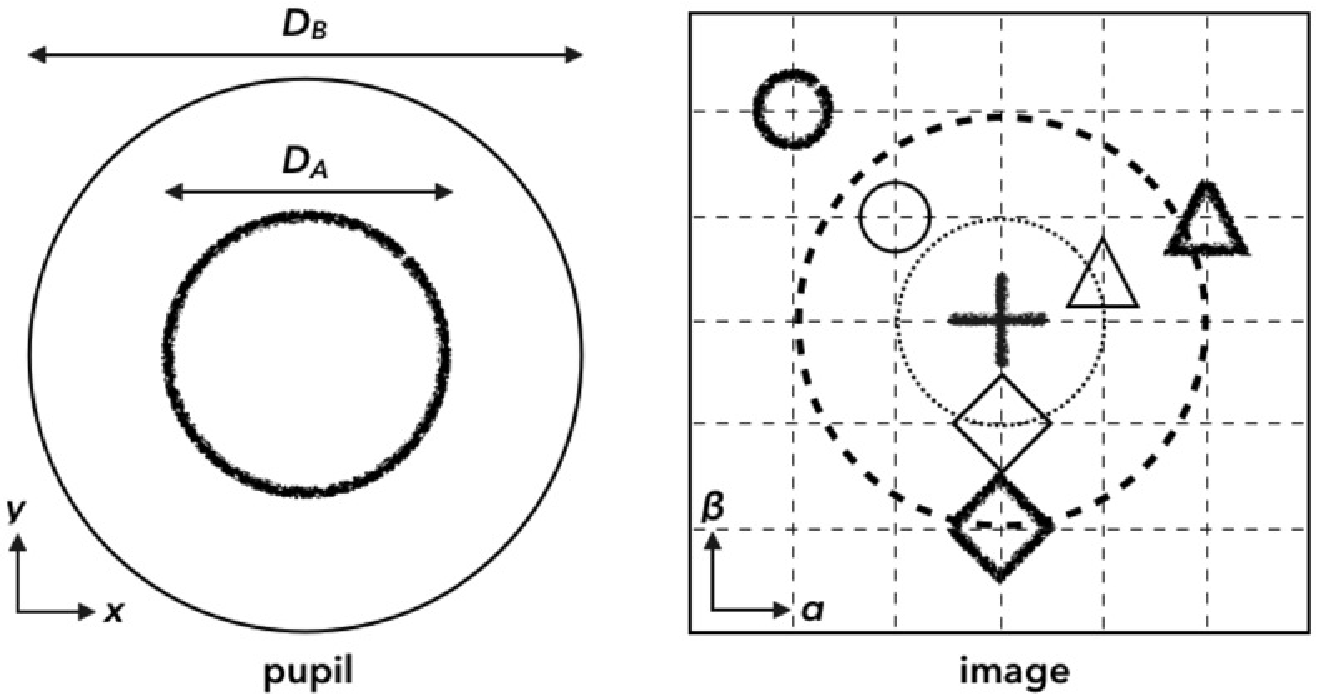}\\
\includegraphics[width=7cm]{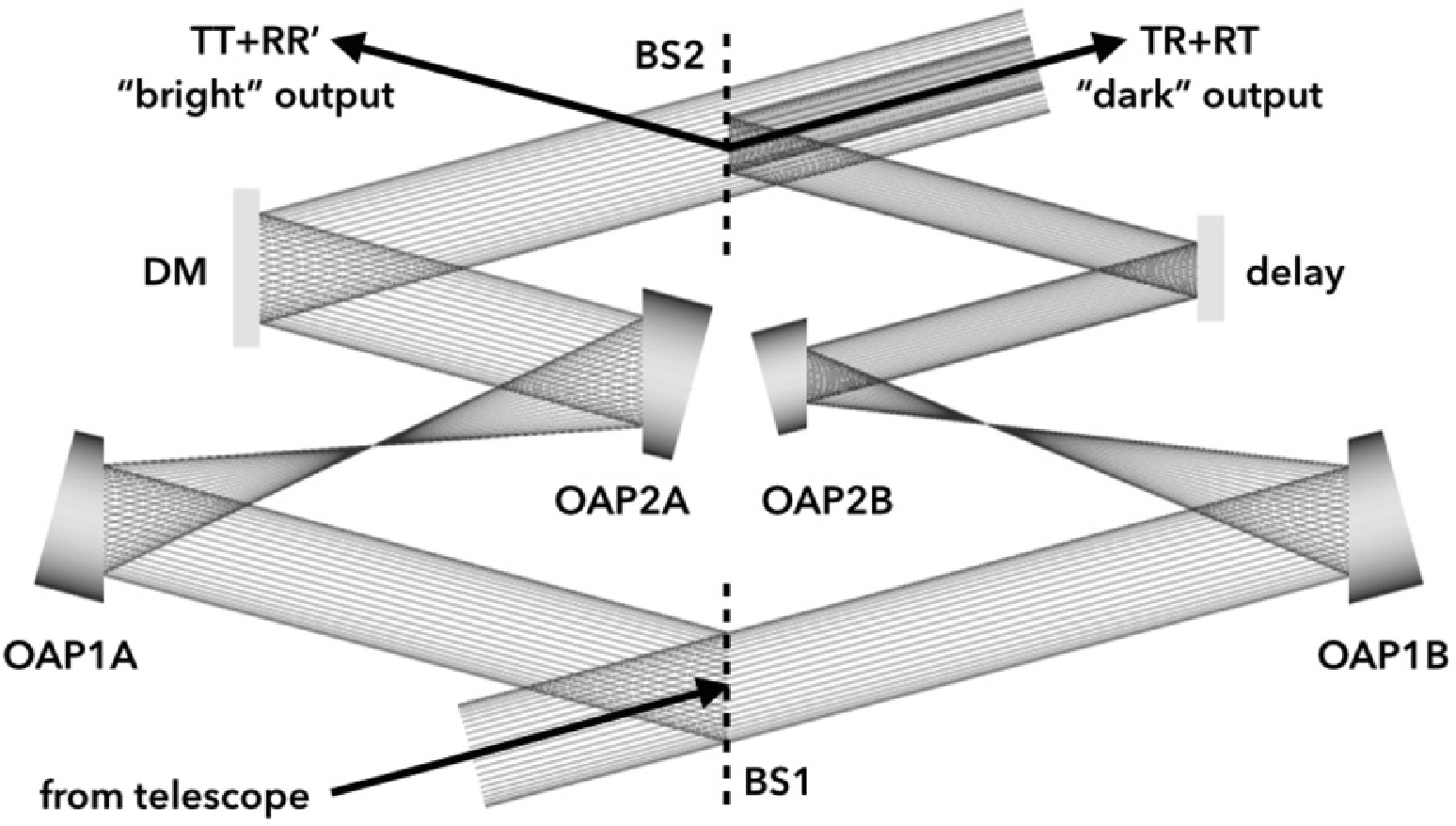}
\caption{A radial shearing interferometer incorporating a DM at a pupil in the 1:1 off-axis parabola (OAP) relay ``A'' arm with corresponding heavy rough lines in the pupil and image footprint insets. The relay in the  ``B'' arm (fine solid lines) introduces 2X angular magnification and a factor of 4X areal beam compression, conserving the radiometric etendue. The dashed lines indicate the approximate Airy disk (2.44$\lambda/D$) size for each arm, and each symbol is a different source center, with the cross being that of the central nulled star.}
\label{fig:schematic2X}
\end{center}
\end{figure}

The PO optics described in this work evolved from the need for full field coverage that requires multiple shears or rolls for the system that has been used to date described in Appendix \ref{appendix:lateral}. The PO approach to nulling is related to the radial shearing interferometer described by \citep{1959H.M.S.O..253,1962AcOpt...9..159H} that could also be adapted for use as a nuller, though with less than adequate performance needed for direct exoplanet detection. Several geometric implementations of the radial shearing interferometer that derive from the original Mach-Zender (MZ), and cyclic or Sagnac interferometers are presented in \citep{1964ApOpt...3..853M}. Here we present a modified MZ layout in Figure \ref{fig:schematic2X} similar to what is shown in Section \ref{sec:formalism}. 

Maintaining symmetry is of key importance to nulling interferometry for reaching high contrast. 
The difference in magnification between the arms of a radial shearing interferometer introduces a field strength asymmetry. 
For the illustrative case presented to scale in Figure \ref{fig:schematic2X}, normalizing to the integrated field within beam A, we have for the bright and dark fringe intensities $I_b=9/4$ and $I_d=1/4$, and the null, which is related to interferometric visibility as
\begin{equation}
N=1-V=1-\frac{I_{b}-I_{d}}{I_{b}+I_{d}}
\end{equation}
would be 0.2. This result falls nine orders of magnitude short of what is needed for directly detecting exoEarths in mature systems. Reducing the difference in magnification to 10\% yields a more favorable 0.0045 null, but this is still far short of what is needed.

The next two-fold limitation of the radial shearing interferometer is 1) the inverse relation between the difference in magnification and the inner working angle (IWA) of the basic radial shearing interferometer used as a nuller, which is the separation angle at which a coronagraph transitions from suppression to transmission, and 2) spreading of the PSF that is the sum of the two beams, which reduces the imaging Strehl ratio and introduces a source of confusion. For the first, it can be intuited that for the Figure \ref{fig:schematic2X} example of 2:1 relative magnification that optical coherence diminishes rapidly as a source moves off-axis, reaching a full large beam diffraction radius separation between focused spots once off-axis by this same angular separation. As mentioned above, the area of non-overlap would be masked out, so the Airy core of the combined beams would have the size of the larger diameter heavy dashed line. The separation should increase linearly and it is conceivable that point source extraction routines could be tailored to recombine the light of the separating spots attributed to each beam. Again reducing to a 10\% difference in magnification, the same separation between points is not reached until five times further off-axis at $10\lambda/D$, with an inverse effect for IWA not occurring until roughly $5\lambda/D$. While the combined PSF is more tolerable out to the edge of this hypothetical nuller's field of view, the IWA would dictate a telescope that is tens of meters in diameter in order to access stellar HZs out to a few tens of parsecs.

Like each of the preceding limitations, leakage attributed to the finite angular extent of the star $\theta$ also scales with the difference in magnification. It is ideal to have a steep a transition from suppression to transmission in order to maintain a small IWA while also minimizing this term, which may be loosely understood as the contribution from a stellar disk filling a coronagraphic dark hole. As a telescope's resolving power increases with it's maximum linear extent (to include long baseline discrete apertures) to separate two distinct sources, it also further increases the degree to which it is sensitive to a star's finite angular extent. In the extreme case of large baselines operating at short wavelengths, the goal is in fact to resolve surface features of stellar disks. For visible observations with a future observatory with a segmented aperture on the scale of 10~m in diameter, Sun-like stars will have diameters ranging from $\theta=0.01-0.1\lambda/D$ out to a distance of $\sim 40$~pc. Completing the discussion of the radial shearing interferometer as a starlight nuller, the stellar extent leakage term for the radial shearing interferometer scales as 
\begin{equation}
L_{\theta}\approx [\theta(m_A/m_B-1)/2]^2
\end{equation}
and again using the same examples of factors of 2 and 10\% difference, this yields $L_{theta}$ on the order of $10^{-4}$ and $5\times 10^{-6}$ for a G2V star at 10~pc observed at 500~nm with a 10~m telescope, which respectively contribute roughly three and two orders of magnitude less leakage than that contributed by field strength imbalance as calculated above.


\begin{thebibliography}{}

\bibitem[Batalha(2014)]{2014PNAS..11112647B} Batalha, N.~M.\ 2014,  Proceedings of the National Academy of Science, 111, 12647 

\bibitem[Brown(1959)]{1959H.M.S.O..253} Brown, D.~S., Interferometry, Symposium No. 11 at National Physical Laboratory, London, England,1959 (H.M. Stationery Office, London, 1960), p. 253.

\bibitem[Brown(2015)]{2015ApJ...799...87B} Brown, R.~A.\ 2015, \apj, 799, 87 

\bibitem[Dumas \& McFee(2014)]{2014SPIE.9281E..09D} Dumas, P., \& McFee, C.\ 2014, \procspie, 9281, 928109 

\bibitem[Guyon et al.(2014)]{2014ApJ...780..171G} Guyon, O., Hinz, P.~M., Cady, E., Belikov, R., \& Martinache, F.\ 2014, \apj, 780, 171

\bibitem[Hariharan \& Sen(1962)]{1962AcOpt...9..159H} Hariharan, P., \& Sen, D.\ 1962, Optica Acta, 9, 159

\bibitem[Hicks et al.(2014)]{2014SPIE.9143E..2SH} Hicks, B.~A., Lyon, R.~G., Bolcar, M.~R., Clampin, M., \& Petrone, P.\ 2014, \procspie, 9143, 91432S 

\bibitem[Krist(2014)]{2014SPIE.9143E..0VK} Krist, J.~E.\ 2014, \procspie, 9143, 91430V

\bibitem[Clampin et al.(2004)]{2004SPIE.5487.1538C} Clampin, M., Melnick,  G.~J., Lyon, R.~G., et al.\ 2004, \procspie, 5487, 1538

\bibitem[Lyon et al.(2008)]{2008SPIE.7010E..45L} Lyon, R.~G., Clampin, M., Melnick, G., et al.\ 2008, \procspie, 7010, 701045 

\bibitem[Lyon et al.(2012)]{2012SPIE.8442E..08L} Lyon, R.~G., Clampin, M., Petrone, P., et al.\ 2012, \procspie, 8442, 844208 

\bibitem[Lyon \& Clampin(2012)]{2012OptEn..51a1002L} Lyon, R.~G., \& Clampin, M.\ 2012, Optical Engineering, 51, 011002 

\bibitem[Lyon et al.(2009)]{2009SPIE.7440E..11L} Lyon, R.~G., Clampin, M., Woodruff, R.~A., et al.\ 2009, \procspie, 7440, 744011

\bibitem[Murty(1964)]{1964ApOpt...3..853M} Murty, M.~V.~R.~K.\ 1964, \ao, 3, 853

\bibitem[Postman et al.(2012)]{2012OptEn..51a1007P} Postman, M., Brown, T., Sembach, K., et al.\ 2012, Optical Engineering, 51, 011007

\bibitem[Serabyn(2000)]{2000SPIE.4006..328S} Serabyn, E.\ 2000, \procspie, 4006, 328

\bibitem[Stark et al.(2014)]{2014ApJ...795..122S} Stark, C.~C., Roberge, A., Mandell, A., \& Robinson, T.~D.\ 2014, \apj, 795, 122

\end{thebibliography}

\clearpage

\end{document}